# Hydrophilization of Liquid Surfaces by Plasma Treatment

Victor Multanen [b], Gilad Chaniel [a,c], Roman Grynyov [a], Ron Yosef Loev [b], Naor Siany [b], Edward Bormashenko [a,b*]

[a] Ariel University, Physics Faculty, 40700, P.O.B. 3, Ariel, Israel
[b] Ariel University, Chemical Engineering and Biotechnology Faculty, 40700, P.O.B. 3, Ariel, Israel
[c] Bar Ilan University, Physics Faculty, 52900, Ramat Gan, Israel

**Abstract**

The impact of the cold radiofrequency air plasma on the surface properties of silicone oils (polydimethylsiloxane) was studied. Silicone oils of various molecular masses were markedly hydrophilized by the cold air plasma treatment. A pronounced decrease of the apparent water contact angles was observed after plasma treatment. A general theoretical approach to the calculation of apparent contact angles is proposed. The treated liquid surfaces demonstrated hydrophobic recovery. The characteristic time of the hydrophobic recovery grew with the molecular mass of the silicone oil.

1. **Introduction.**

Plasma treatment (low and atmospheric-pressure) is broadly used for the modification of surface properties of solid polymer materials [1-6]. The plasma treatment creates a complex mixture of surface functionalities which influence surface physical and chemical properties and results in a dramatic change in the wetting behaviour of the surface [7-16]. Not only the chemical structure but also the roughness of the surface is affected by the plasma treatment, which also could change the wettability of the surface [17]. Plasma treatment usually strengthens the hydrophilicity of treated solid polymer surfaces. However, the surface hydrophilicity created by plasma treatment is often lost over time. This effect of decreasing hydrophobicity is called "hydrophobic recovery" [18-28]. The phenomenon of hydrophobic recovery is usually attributed to a variety of physical and chemical processes, including: 1) re-arrangement of chemical groups of the surface exposed to plasma treatment, due to the conformational mobility of polymer chains; 2) oxidation and degradation reactions at the plasma treated surfaces; 3) diffusion of low molecular weight products from the outer layers into the bulk of the polymer, 4) plasma-treatment induced diffusion of additives introduced into the polymer from its bulk towards its surface [23]. Occhiello *et al*. classified the complicated processes occurring under hydrophobic recovery according to their spatial range, i.e. short-range motions within the plasma-modified layer, burying polar groups away from the surface and long-range motions, including diffusion of non-modified macromolecules or segments from the bulk to the surface [27]. A phenomenological model of hydrophobic recovery has been proposed recently by Mortazavi and Nosonovsky [25].

The research in the field has been focused on the influence exerted by plasma on solid surfaces. Our work is the first that reports the impact of cold radiofrequency plasma on liquid polymer surfaces (silicone oils of various molecular masses). The interest in wetting regimes inherent to water/oils pairs was boosted in the last decade due to the extremely low contact hysteresis demonstrated by these systems [29-33].

**2. Experimental.**

Aluminium foil was coated with honeycomb polycarbonate (PC) films, by the fast dip-coating process. As a result, we obtained typical "breath-figures" self-assembly patterns, depicted in Fig. 1. Honeycomb PC coating was obtained according to the protocol described in detail in Refs. 34-35. The average radius of pores was about 1.5 μm. The average depth of pores as established by AFM was about 1 μm.

Polydimethylsiloxane (PDMS) oils with molecular masses of 5600, 17500 and 24000 g/mole, denoted for brevity in the text respectively as PDMS1, PDMS2 and PDMS3 were supplied by Aldrich. PC porous coatings were impregnated by all kinds of aforementioned PDMS oils. The thickness of PDMS oil layers was established by weighing as 20±2 μm.

Liquid PDMS layers were exposed for 30 s to a radiofrequency (13MHz) inductive air plasma discharge under the following parameters: pressure 133 Pa, power 18 W, ambient temperature. The details of the experimental setup are supplied in Ref. 16.

Apparent contact angles were measured by a Ramé-Hart Advanced Goniometer (Model 500-F1). A series of 3 experiments was carried out for every aforementioned PDMS oils. For the study of the hydrophobic recovery, the apparent contact angles were measured every few minutes during the first 2 hours after the plasma treatment.

3. **Results and discussion.**

   **3.1 Influence of plasma treatment on apparent contact angles of PDMS.**

It is plausible to suggest that the cold plasma treatment increased the surface energy of silicone liquids [13, 36]. The natural measure of the surface energy is an apparent contact angle, so-called APCA [37-39]. However, in our case the accurate establishment of the APCA is far from trivial. When a water droplet is placed on the silicone oil, it is coated with the oil, as depicted in Fig. 3. This wetting situation is well explained by the analysis of the spreading parameter $S$ governing the wetting situation [37-39]:

$$S = \gamma - (\gamma_{oil} + \gamma_{oil/water}),  \quad (1)$$

where $\gamma = 70\frac{mJ}{m^2}, \gamma_{oil} = 20\frac{mJ}{m^2}, \gamma_{oil/water} = 23-24\frac{mJ}{m^2}$ are interfacial tensions at water/vapor, oil/vapor and oil/water interfaces respectively (interfacial tensions are taken from the literature data [40]). Substituting the aforementioned values in Exp. 1, we obtain $S > 0$; in this case, the silicone oil is expected to coat the water droplet [41-43]. The formation of the silicone oil layer coating the droplet was observed experimentally. For the purposes of visualization the orange pigment 4-(Phenyldiazenyl)benzenamine (Acid yellow 9), soluble in PDMS but insoluble in water, was added to PDMS oils. The water droplet placed on the liquid PDMS layer, impregnating the porous PC, was painted in orange colour, as shown in Fig. 4.

Thus, the actual wetting regime displayed schematically in Fig. 3, illustrated with Fig. 4, and discussed recently in detail in Refs. 41-43, is rather intricate. The situation is complicated by the "wetting ridges" formed in the vicinity of the contact line, discussed in Refs. 41-43 and shown schematically in Fig. 3. The complicated shape of the water/vapor interface, exhibiting a flex point, is noteworthy, making an accurate measurement and interpretation of the contact angle quite challenging. For the analysis of the wetting situation, we implemented the following experimental procedure. We established that the thinner silicone oil layers resulted in smaller wetting ridges. Thus we coated porous PC substrates with silicone layers having a thickness of about 20μm, resulting in negligible wetting ridges and allowing accurate building of a tangent line within a traditional goniometric procedure.

The initial "as placed" APCAs (see Refs. 44-45) of plasma-treated PDMS layers, taken immediately after the exposure of silicone oils to plasma and the "as placed" APCA of non-treated films are supplied in Table 1. It is clearly seen that APCAs decreased markedly after the plasma treatment, as shown in Fig. 5. Thus, we conclude that the cold air plasma treatment significantly increased the surface energies of polymer liquids; i.e. the surfaces of silicone oils were hydrophilized by the plasma treatment. It is noteworthy that the experimental scattering of APCAs was low.

This hydrophilization, however, does not persist forever. The APCAs grew with time. This effect is called "hydrophobic recovery" [21-25, 28]. The kinetics of hydrophobic recovery is illustrated with Fig. 6.

The time dependencies of the APCA were approximated by the empirical formula [25]:

$$\theta^*(t) = \tilde{\theta}(1 - e^{-\frac{t}{\tau}}) + \theta_0 = \theta_{sat} - \tilde{\theta}e^{-\frac{t}{\tau}}, \qquad (2)$$

where $\theta_0$ corresponds to the initial APCA established immediately after the plasma treatment, $\tau$ is the characteristic time of restoring the contact angle, $\tilde{\theta}$ is the fitting parameter, and $\theta_{sat} = \tilde{\theta} + \theta_0$ corresponds to the saturation contact angle, as calculated by fitting of the experimental data with Exp. 2. The characteristic times of the hydrophobic recovery for various PDMS oils calculated with Exp. 2 are summarized in Table 2. It is seen that these times are on the order of magnitude of 4-60 minutes.

It seems plausible to relate the pronounced hydrophilization of PDMS oils at least partially to the re-orientation of hydrophilic groups of PDMS towards the PDMS/air interface. This effect disappears with time, due to the thermal agitation of hydrophilic groups, explaining the well-known process of hydrophobic recovery [23, 25]. The idea that the contact angle of a water droplet on the surface of a solid polymer depends on whether the hydrophilic moiety of the polymer molecule is oriented towards the solid/air interface or towards the bulk of the solid, and not only on the hydrophilicity of the molecule, has been put forward in the classic work carried out by Yasuda *et al*. [46]. It is reasonable to suggest that a similar situation occurs for polymer liquids.

Shorter polymer chains have less moment inertia of rotation; longer polymer chains will unfold more slowly. Evidently, this results in the characteristic time of the hydrophobic recovery $\tau$ increasing with the molecular mass [47]. Consider the value of the characteristic time of the hydrophobic recovery, which is $\tau \cong 4 - 60 \min$. It seems instructive to compare the kinetics of hydrophobic recovery observed in our experiments with the kinetics of the variation of adhesion force at liquid/solid interfaces reported recently by Tadmor *et al*. [48-49]. Tadmor *et al*. studied the temporal change of the adhesion force at the solid/liquid interface with the centrifugal adhesion balance for the hexadecane/Teflon system. They reported the characteristic time of this change on the order of magnitude of 10 min, coinciding quantitatively with our findings. Tadmor *et al*. related the change in the adhesion force to molecular reorientation of polymer chains which admittedly also takes place in our experiments [48-49]. It should be emphasized that the characteristic times of hydrophobic recovery established for polymer liquids in our experiments are much shorter than those typical for solid polymers [18-28]. It seems plausible to relate this observation to the higher mobility of polymer chains in the liquid state.

### 3.2 Interpretation of the apparent contact angles for water/liquid PDMS systems.

As it was already mentioned, the interpretation of APCAs in the wetting situation depicted in Figs. 3-4, when liquid PDMS coats a water droplet, is far from trivial. The most general thermodynamic approach to the establishment of the APCAs is the implementation of the transversality conditions of the appropriate variational problem of wetting [39, 50-52]. When a water droplet is deposited on the silicone layer as described in Fig. 3, its free energy $G$ (the energy of the "wetting ridge" is neglected) could be written as :

$$G[h(x,y)] = \iint_S \left[ (\gamma_{oil} + \gamma_{oil/water} + P(e))\sqrt{1+(\nabla h)^2} + (\gamma_{oil/water} - \gamma_{oil}) \right] dxdy, \quad (3)$$

Where $\gamma_{oil}$ is the oil/air interfacial tension, $e$ is the thickness of the PDMS liquid layer coating a water droplet (see Fig. 3), $h(x,y)$ is the local height of the liquid surface above the point $(x,y)$ of the substrate (it is supposed latently that there is no difference between surface tensions and surface energies for $\gamma_{oil}$, $\gamma_{oil/water}$), and the integral is extended over the wetted substrate area. The first term of the integrand presents the capillary energy of the liquid "composite" cap, and the second term describes the change in the energy of the lower PDMS layer covered by water. $P(e)$ is the term resulting from the disjoining pressure $\Pi(e)$ (see Refs. 37-39, 53).

$$\Pi(e) = -\frac{dP}{de} \qquad (4)$$

Gravity is neglected, because it has no influence on the APCA [39, 50-51]. If we restrict ourselves with an axially symmetric "composite" droplet (shown in Fig. 3), the free energy $G$ is given by the following expression:

$$G(h,h') = \int_o^a \left[2\pi(\gamma_{oil} + \gamma_{oil/water} + P(e))x\sqrt{1+h'^2} + 2\pi x(\gamma_{oil/water} - \gamma_{oil})\right]dx, \qquad (5)$$

where $a$ is the contact radius of the droplet (see Fig. 3). We also suppose that the PDMS coated water droplet does not evaporate (this assumption is well justified, because PDMS actually does not evaporate); thus the condition of the constant volume $V$ should be considered as:

$$V = \int_o^a 2\pi x h(x) dx = \text{const.} \qquad (6)$$

Thus the problem is reduced to the minimization of the functional:

$$G(h,h') = \int_0^a \tilde{G}(h,h',x)dx, \qquad (7a)$$

$$\tilde{G}(h,h',x) = 2\pi(\gamma_{oil} + \gamma_{oil/water} + P(e))x\sqrt{1+h'^2} + 2\pi x(\gamma_{oil/water} - \gamma_{oil}) + 2\pi\lambda x h. \qquad (7b)$$

If the endpoints of the contact line are free to move, the transversality condition at the endpoint $a$ yields [52]:

$$(\tilde{G} - h'\tilde{G}'_{h'})_{x=a} = 0, \qquad (8)$$

where $\tilde{G}'_{h'}$ denotes the $h'$ derivative of $\tilde{G}$. Minimization of the Functional (7a-b) according to the procedure described in detail in Refs. 39, 50-51 yields for the apparent contact angle:

$$\cos\theta^* = \frac{\gamma_{oil} - \gamma_{oil/water}}{\gamma_{oil} + \gamma_{oil/water} + P(e)}. \qquad (9)$$

Calculation of $P(e)$ depends on the specific function representing the peculiar Derjaguin isotherm $\Pi(e)$, inherent for water/silicone oil system [38, 53-54]. When the effect of disjoining pressure is neglected, we obtain the quite predictable expression for APCA:

$$\cos\theta^* = \frac{\gamma_{oil} - \gamma_{oil/water}}{\gamma_{oil} + \gamma_{oil/water}} \quad , \qquad (10)$$

which may be easily interpreted in terms of the equilibrium of interfacial forces acting on the unit length of the contact line (which is not the "triple" line in our case, due to the composite nature of the liquid cap). Let us check the applicability of Exp. 10 for prediction of APCAs. When we substitute the aforementioned values of $\gamma_{oil}, \gamma_{oil/water}$ in Exp. 10, we obtain for the value of APCA $\theta^* = 94 - 95^0$ in satisfactory agreement with the experimentally observed $\theta^*_{exp} = 89 - 103^0$ (see Table 1). It means that the disjoining pressure $P(e)$, neglected in Exp. 10, does not play a crucial role in the constituting APCA.

Plasma treatment influences the $\gamma_{oil/water}$, resulting in a dramatic change of APCA ($\theta^*$). It is noteworthy that a more accurate calculation of $\theta^*$ according to Exp. 9 turns out to be a challenging task, due to the fact that the thickness of the PDMS oil layer coating a water cap and the precise expression describing the Derjaguin isotherm $\Pi(e)$ are not well-known [37-39, 53-54].

## 4. Conclusions

We conclude that the cold radiofrequency air plasma treatment of organic liquids (polydimethylsiloxane) markedly modified (increased) their surface energy. This modification resulted in the pronounced hydrophilization of the polymer liquid/air interface. The observed hydrophilization was followed by hydrophobic recovery. The characteristic time of the hydrophobic recovery, which is on the order of magnitude of dozens of minutes, grew with the molecular mass of polydimethylsiloxane. The wetting regime of the polydimethylsiloxane oil surface was rather complicated: a water droplet placed on the silicone oil was coated with it. A theoretical approach to the interpretation of experimentally observed contact angles is proposed. The developed approach is based on the analysis of a variational problem of wetting in the system water/silicone oil. The apparent contact angle, observed experimentally, coincides well with the value resulting from the application of the transversality conditions of the variational problem of wetting. We demonstrated that not only solids but also liquids may be hydrophilized effectively by the cold radiofrequency plasma treatment.

**Acknowledgements**

We are thankful to Mrs. Y. Bormashenko for her help in preparing this manuscript. The Authors are thankful to Mrs. N. Litvak for SEM imaging of PC substrates.
**References**

[1]  H. K. Yasuda, J. Wiley & Sons, Plasma polymerization and plasma treatment, New York, 1984.

[2]  M. Strobel, C. S. Lyons, K. L. Mittal (Eds), Plasma surface modification of polymers: relevance to adhesion, VSP, Zeist, The Netherlands, 1994.

[3]  M. Thomas, K. L. Mittal (Eds), Atmospheric Pressure Plasma Treatment of Polymers, Plasma surface modification of polymers: relevance to adhesion, Scrivener Publishing, Wiley, Beverly, USA, 2013.

[4]  M. Lehocky, H. Drnovska, B. Lapcikova, A. M. Barros-Timmons, T. Trindade, M. Zembala, L. Lapcik Jr., Plasma surface modification of polyethylene, Colloids Surf., A 222 (2003) 125-131.

[5]  E. C. Preedy, E. Brousseau, S. L. Evans, S. Perni, P. Prokopovich, Adhesive forces and surface properties of cold gas plasma treated UHMWPE, Colloids Surf., A 2014, http://dx.doi.org/10.1016/j.colsurfa.2014.03.052

[6]  R. M. Cámara, E. Crespo, R. Portela, S. Suárezd, L. Bautista, F. Gutiérrez-Martín, B. Sánchez, Enhanced photocatalytic activity of $TiO_2$ thin films on plasma-pretreated organic polymers, Catalysis Today 230 (2014) 145–151.

[7]  E. Occhiello, M. Morra, F. Garbassi, D. Johnson, P. Humphrey, SSIMS studies of hydrophobic recovery: Oxygen plasma treated PS Appl. Surf. Sci. 47 (1991) 235-242.

[8]  R. M. France, R. D. Short, Plasma treatment of polymers: The Effects of energy transfer from an Argon plasma on the surface chemistry of polystyrene, and polypropylene. A High-Energy resolution X-ray photoelectron spectroscopy study, Langmuir 14 (1998) 4827–4835.

[9]  R. M. France, R. D. Short, Plasma treatment of polymers: Effects of energy transfer from an argon plasma on the surface chemistry of poly(styrene), low density poly(ethylene), poly(propylene) and poly(ethylene terephthalate), J. Chem. Soc., Faraday Trans. 93 (1997) 3173-3178.

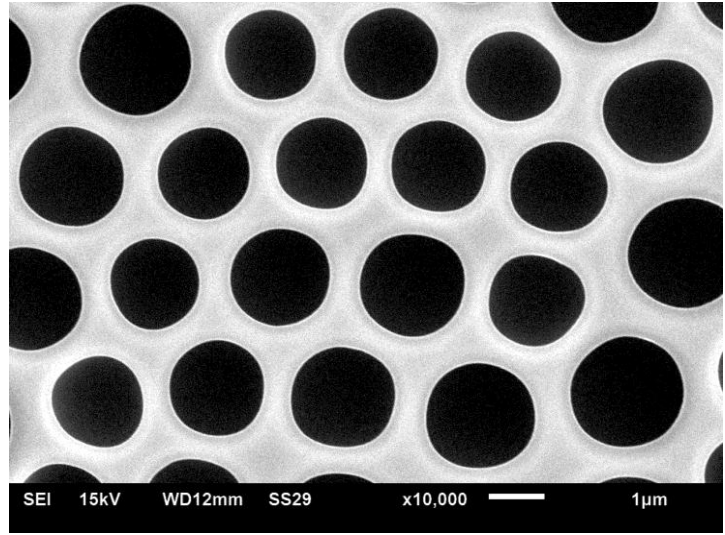

Fig. 1. Polycarbonate honeycomb coating of Al foil obtained with "breath-figures", carried out in a humid atmosphere. Scale bar is 1 μm.

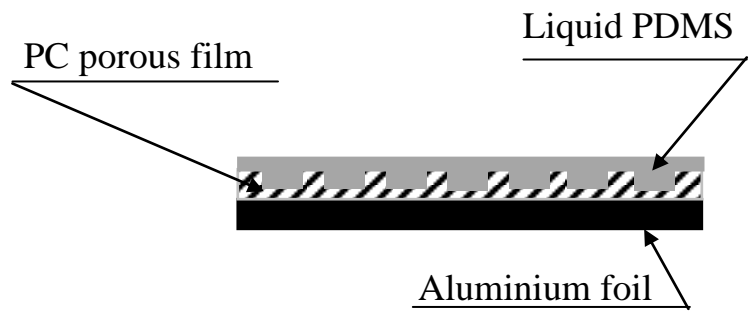

Fig 2. Scheme of PDMS oil layers exposed to air plasma discharge.

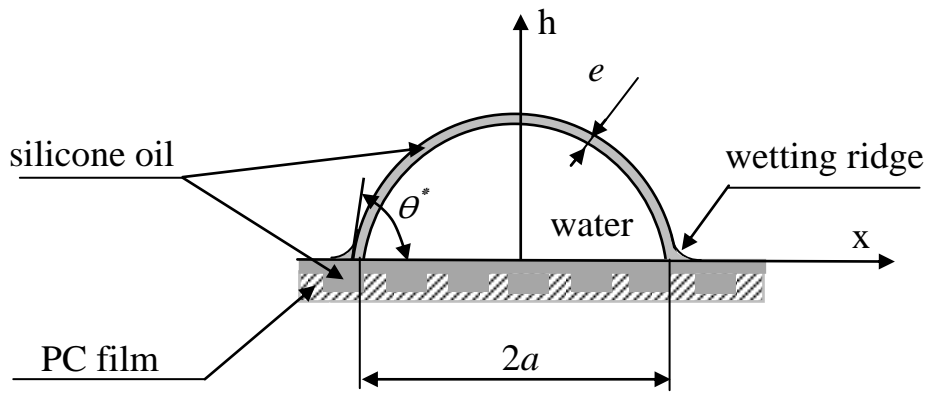

Fig. 3. Wetting regime for water droplets deposited on the silicone oil layer. $\theta^*$ is the apparent contact angle.

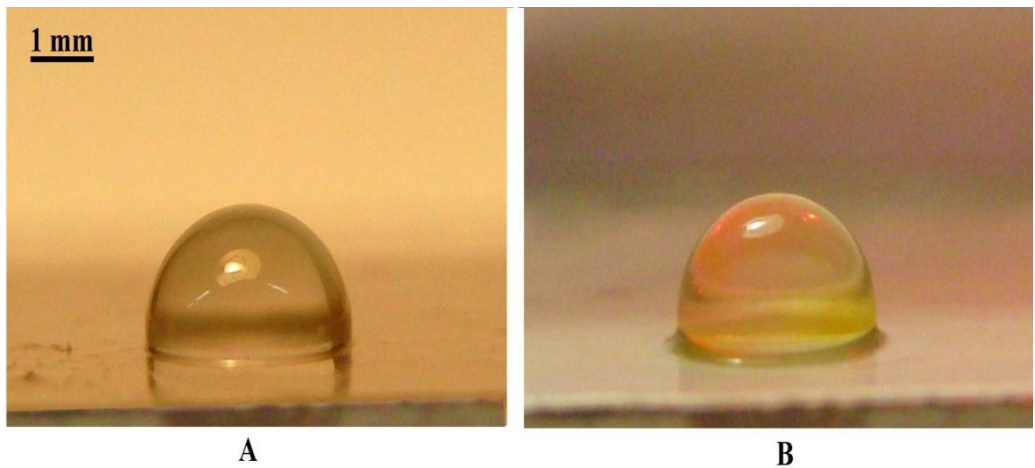

Fig. 4. A. 8 µl water droplet coated with a layer of the non-colored silicone oil. B. 8 µl water droplet coated with a layer of the silicone oil, colored with the orange pigment.

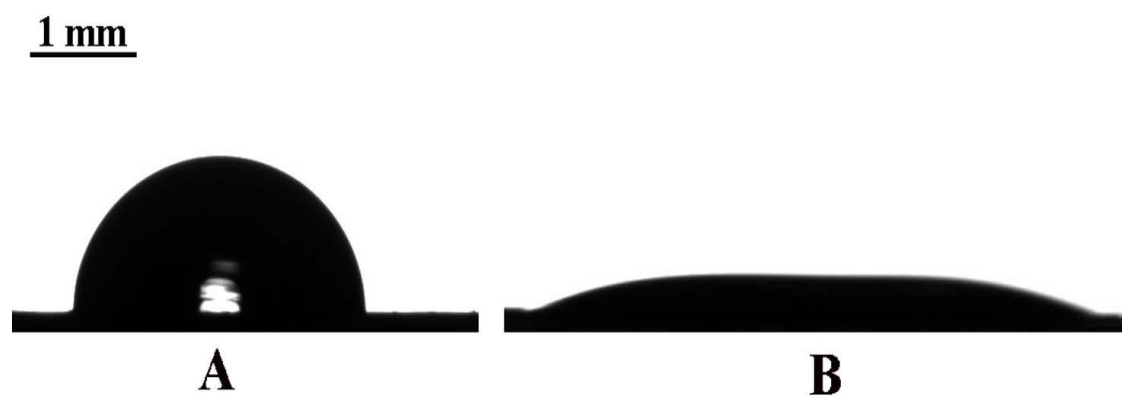

Fig. 5. 8 µl water droplet deposited on non-treated (A) and plasma-treated (B) PDMS1 silicone oil.

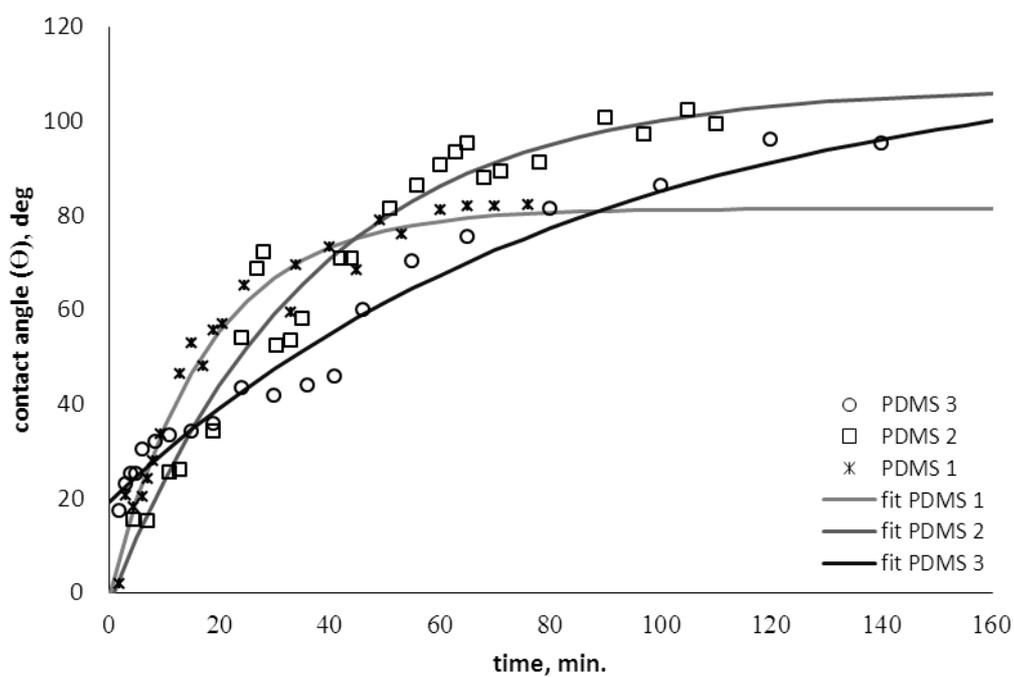

Fig. 6. The apparent contact angle (APCA) $\theta^*$ as a function of the time, $t$. Solid lines depict the exponential fitting of experimental data with Exp. 2.